\documentclass[11pt]{article}

\usepackage{graphicx}
\usepackage{color}
\usepackage[hypertex]{hyperref}
\usepackage{a4wide}

\title{Modelling the Navigation Potential of a Web Page}

\author{Trevor Fenner, Mark Levene and George Loizou \\
Birkbeck, University of London \\
London WC1E 7HX, U.K. \\ \{trevor,mark,george\}@dcs.bbk.ac.uk}

\date{}

\begin{document}

\maketitle

\newtheorem{theorem}{Theorem}[section]
\newtheorem{corollary}[theorem]{Corollary}
\newtheorem{lemma}[theorem]{Lemma}
\newtheorem{proposition}[theorem]{Proposition}
\newtheorem{definition}{Definition}[section]
\newtheorem{algorithm}{Algorithm}
\newtheorem{example}{Example}[section]

\begin{abstract}

Suppose that you are navigating in ``hyperspace'' and you have reached a web page with
several outgoing links you could choose to follow. Which link should you choose in such an online scenario? One extreme case is that you know exactly where you are heading and you have no problem in
choosing a link to follow. In all other cases, when you are not sure where the
information you require resides, you will initiate a navigation (or ``surfing'')
session. This involves pruning (or discounting) some of the links and following one of
the others, where more pruning is likely to happen the deeper you navigate. In terms of
decision making, the utility of navigation diminishes with distance until finally the
utility drops to zero and the session is terminated. Under this model of navigation, we
call the number of nodes  that are available after pruning, for browsing within a
session,  the {\em potential gain} of the starting web page. Thus the parameters that
effect the potential gain are the local branching factor with respect to the starting
web page and the discount factor.

\smallskip

We first consider the case when the discounting factor is geometric. We show that the distribution
of the effective number of links that the user can follow at each navigation step after pruning,
i.e. the number of nodes added to the potential gain at that step, is given by the {\em erf}
function, which is related to the probability density function for the Normal distribution. We
derive an approximation to the potential gain of a web page and show that this is numerically a very accurate estimate. We also obtain lower and upper bounds on the potential gain.
We then consider a harmonic discounting factor and show that, in this case, the potential gain at each step is closely related to the probability density function for the Poisson distribution.

\smallskip

The potential gain has been applied to web navigation where, given no other information,
it helps the user to choose a good starting point for initiating a surfing session.
Another application is in social network analysis, where the potential gain could
provide a novel measure of centrality.

%The other extreme is that you are completely lost and as far as you are
%concerned any link is as good as the other. In this case you should
%choose the link which leads to the largest subgraph within the horizon of
%your search, the horizon being determined by the number of clicks you
%will employ before stopping. The reason for this being that, everything
%else being equal, you should navigate through the densest region of the
%space. In the terminology of this work you should follow the link leading
%to a page with the largest {\em potential gain}. The middle ground is the
%situation when you are not completely lost but are still unsure which
%link to follow. In this case the best policy seems to be some weighting
%of your beliefs and the potential gain.

\end{abstract}

\section{Introduction}

In order to find information on the World-Wide-Web, ``surfers'' often adopt the following two-stage
strategy \cite{LEVE05}. First they submit their query to a global web search engine, such as Google
or Yahoo, which directs them to the home page of the subdomain within the web site that is likely to contain the information they are looking for. Then they navigate within this web site by following
hyperlinks until they either find the information they are seeking, or they restart their search by
reformulating their original query and then repeating the process. In some cases users simply give
up their search task when they lose the context in which they were browsing and are unsure how to
proceed in order to satisfy their original goals. This phenomenon is known as the {\em navigation
problem} \cite{LEVE00a} or colloquially as ``getting lost in hyperspace'' \cite{NIEL00}.

\medskip

Although, as far as we know, global web search engines attach higher weights to home pages
than to other pages, they do not have a general mechanism to take into consideration the
navigation potential of web pages. Our aim in this paper is to investigate the problem of
finding ``good'' starting points for web navigation that are independent of the user's
query. Once we have available such a measure, we can weight this information into the
user's query in order to find ``good'' points for starting navigation given the actual
query. Hereafter we shall refer to the measure of navigation potential of a web page as
its {\em potential gain}. We note that the application that initially led us to look into
the potential gain is the search and navigation engine we have developed for
semi-automating user navigation within web sites \cite{WHEE03}, but we believe that this
notion has wider applicability within the general context of web search tools
\cite{LEVE04a}.

\medskip

In view of the above, we would like to choose a web page (or more technically a URL, i.e. a {\em
Uniform Resource Locator}) from which to start navigation that in some well-defined sense maximises
the potential of the user to realise his/her ``surfing'' goal. The only a priori information that may be
available is partial knowledge of the topology of the web, i.e. the set of URLs which are reachable
from a given starting URL. This information amounts to some knowledge about the density of web pages
in the neighbourhood of the starting URL. Essentially, if this neighbourhood is denser, i.e. we can
potentially reach many URLs in a short distance, then we consider the potential gain, or utility, of
this URL to be high. For example, the home page of a web site is normally a ``good'' starting URL
for navigation precisely for the reason that there is a wealth of information reachable from it.

\medskip

Assuming that we are navigating within the web graph, the potential gain of a starting
URL is, informally, the number of URLs that can be reached from the starting point,
where at each step the number of outgoing links is successively discounted depending on the distance
from the starting point. We investigate two discounting functions, geometric and
harmonic. For geometric discounting we show that the potential gain values follow a
Normal distribution with respect to the distance from the starting point, while for
harmonic discounting the distribution is Poisson. Moreover, for geometric discounting,
we derive an approximation to the potential gain, which is numerically very accurate,
and also derive lower and upper bounds.

\medskip

The rest of the paper is organised as follows. In Section~\ref{sec:pg-node} we give a formal definition of the
potential gain of a web page, and derive bounds on it, assuming a geometric
discounting factor. In Section~\ref{sec:analysis} we provide a brief computational analysis of the
distribution of the potential gain values and demonstrate the tightness of the derived bounds. In
Section~\ref{sec:poisson} we investigate the potential gain when utilising a harmonic discounting factor.
Finally, in Section~\ref{sec:concluding} we give our concluding remarks. For graph-theoretic
concepts and background we refer the reader to \cite{BUCK90}.

\section{The Potential Gain of a Web Page}
\label{sec:pg-node}

%Many researchers in the hypertext and information visualisation
%communities suggest that to navigate effectively and efficiently without
%getting lost, requires readers to be aware of their location in the
%information space and to be able to pick up the "scent" \cite{PIRO97} of
%what their next destination might be and then follow the right trail
%leading to this destination.

Let us assume that the user is in the midst of a navigation session having started from
a certain URL, say $U$. The user is browsing a web page and has to decide whether to follow
one of the links on the page or to terminate the session. We make the assumption that
{\em the utility of browsing a web page diminishes with the distance of that page from
the starting URL} $U$. This assumption is consistent with experiments carried out on
real web data \cite{HUBE98,LEVE00b}. So a user browsing a page at distance $d$ from
$U$ will prune from the links actually present those considered to be not worth following;
and for larger $d$ a larger proportion of the links will be pruned. For this purpose
we define the {\em discount factor} $\delta$, with $0 < \delta < 1$, and assume that, at
distance $d$, the user will only inspect the fraction $\delta^{d}$ of the currently available links,
prior to following one of these. Some of the links may be pruned because they lead to pages that the
user has already inspected, whilst others may be pruned as a result of filtering, for
example, by picking up the ``scent of information'' \cite{PIRO97}.

\medskip

We model the web graph as a directed graph $G = (\cal{U}, \cal{E})$ having a set of
nodes (or URLs) $\cal{U}$ and a set of arcs (or links) $\cal{E}$. For
convenience we will assume that $G$ is strongly connected, although this restriction
could be relaxed. To formalise our model of the user, we need to estimate the {\em local
branching factor} $\beta$ of $G$ with respect to a given starting URL $U$: this is a local
estimate of the number of outlinks per node.
For this purpose we define an integer parameter $\Delta$, called {\em clicks},
where $\Delta \ge 1$; this denotes the mean number of clicks (rounded down) a user makes during a
navigation session, i.e. links she follows before terminating her session.
(See \cite{HUBE98,LEVE00b} for an analysis of the distribution of {\em clicks}.)
The local branching factor gives an estimate of how many
links, on average, the user has to choose from, and {\em clicks} gives an estimate of the
number of links, on average, she will traverse during a navigation session. Given
$\Delta$, let $reach(U)$ be the subgraph of $G$ induced by traversing $G$ in a breadth-first manner to depth $\Delta$, starting from $U$. We then define $\beta$ as the average branching factor (i.e. out-degree) of the nodes in $reach(U)$. (We note that, in our breadth-first traversal, we do not keep a record of the nodes visited, so we may visit a node more than once.)  In an online scenario an estimate of $\beta$ may be obtained by sampling in the vicinity of $U$, or from preprocessed log data of previous surfers who have visited $U$. Suppose we have determined the structure of the subgraph of  $reach(U)$ obtained by searching to some depth $\Delta^* \le \Delta$. We can then compute $\beta_d$, the average branching factor of the nodes at depth $d$, $0 \le d < \Delta^*$, as the arithmetic mean of the branching factors of the nodes at depth $d$. In order to maintain consistency with the total number of nodes at level $\Delta^*$, we suggest using the geometric mean of the $\beta_d$, $0 \le d < \Delta^*$, as an estimate of $\beta$.
An estimate of $\delta$ can then be obtained from $\beta$ and $\Delta$, as we show later.

\smallskip

Hence, given $\beta$, the {\em effective branching factor} at depth $i$ is $\beta \delta^i$, and the potential number of available nodes at this depth is approximately
\begin{equation}\label{eq:pg-depthi}
\beta \delta^0 \ \beta \delta^1 \cdots \beta \delta^{i-1} =
\beta^i \ \delta^{i (i-1) / 2}.
\end{equation}
\medskip

The {\em total potential gain} of $U$, denoted by $PG(U)$, is simply the total number of
available nodes at all depths, i.e.
\begin{equation}\label{eq:pg-define}
PG(U) = \sum_{i=0}^\infty \beta^i \ \delta^{i (i-1) / 2}.
\end{equation}
\smallskip

We observe that the potential gain, as defined in the above equation, differs from the PageRank
\cite{BIAN04,BERK05} -- the most studied link analysis metric -- in that the discounting factor
$\delta$ gives rise to a double exponential, thus guaranteeing that the effective branching factor
monotonically decreases to zero. Consequently, the portion of the web graph that is potentially
reachable during a session is bounded. In the PageRank model, the effective branching factor is
always greater than one and, consequently, the PageRank depends on the entire web graph. Moreover,
in the PageRank model, the (random) surfer wanders on ad infinitum, whereas, in the navigation-based
model presented here, the length of the surfer's session is limited by the diminishing branching
factor. This allows us to approximate (\ref{eq:pg-define}) using the {\em erf} function.
(We note that the potential gain may be viewed as a generalised ranking algorithm \cite{BAEZ06}
with a double exponential damping function; this type of damping function was not considered in \cite{BAEZ06}.)

\medskip

Setting $a = \beta \delta^{- 1/2}$, $\theta = \delta^{1/2}$ and $\lambda^2 = \ln (1
/ \theta)$, the potential gain of $U$ up to depth $d$, denoted by
$PG_d(U)$, is given by
\begin{equation}\label{eq:pg}
PG_d(U) = \sum_{i=0}^d a^i \ \theta^{i^2} = \sum_{i=0}^d a^i \ e^{- \lambda^2 i^2}.
\end{equation}
\smallskip

To approximate $PG(U)$, we need to find the greatest depth $d$ such that
\begin{displaymath}
a^d \ \theta^{d^2} \ge 1, \quad {\rm i.e.} \quad a \theta^d \ge 1,
\end{displaymath}
since for greater depths the number of available nodes will be less than one; this value of $d$ corresponds to $\Delta$. Thus
\begin{displaymath}\label{eq:depth}
\Delta = \left\lfloor \frac{\ln (a)}{\ln (1 / \theta)} \right\rfloor =  \left\lfloor \frac{\ln
(a)}{\lambda^2} \right\rfloor.
\end{displaymath}
\medskip

Now, let
\begin{equation}\label{eq:N}
N = \frac{\ln (a)}{\lambda^2} = \frac{2 \ln \beta}{\ln (1 / \delta )} + 1,
\end{equation}
noting that $\Delta = \lfloor N \rfloor$. (Since $\Delta \approx N$, given $\beta$ and $\Delta$, we can thus derive an approximation to $\delta$.)

\medskip

We claim that $a^x \theta^{x^2}$ attains its maximum at $x = N / 2$. To show this we take its
derivative, obtaining
\begin{displaymath}\label{eq:reach-derivative}
\frac{\mathrm{d}}{\mathrm{d} x} \ \bigg( a^x \theta^{x^2} \bigg) = a^x \theta^{x^2}
\bigg( \ln(a) + 2 x \ln(\theta) \bigg),
\end{displaymath}
which is equal to zero at
\begin{equation}\label{eq:max-total-leaves}
x = \frac{\ln (a)}{2 \ln (1 / \theta)} =  \frac{\ln (a)}{2 \lambda^2} =
\frac{N}{2}.
\end{equation}
\smallskip

It can be verified that the second derivative of $a^x \theta^{x^2}$ at $x = N/2$ is
negative and thus this function has a maximum at this point.

\medskip

We next proceed to find an approximation of (\ref{eq:pg-define}) by using the
Euler-Maclaurin summation formula \cite{FROB65}.

\smallskip

Now $\theta = e^{- \lambda^2}$ and, from (\ref{eq:N}), $a = e^{\lambda^2 N}$, so
\begin{displaymath}\label{eq:exp1}
a^i \theta^{i^2} = e^{\lambda^2 (N i - i^2)} = e^{\lambda^2 N^2 / 4} \ e^{- \lambda^2 (i -
N/2)^2}.
\end{displaymath}

Therefore, from (\ref{eq:pg}),
\begin{equation}
PG_d(U) = e^{\lambda^2 N^2 / 4} \ \sum_{i=0}^d e^{- \lambda^2 (i - N/2)^2} =
e^{\lambda^2 N^2 / 4} \ \sum_{i=0}^d f(i),
\end{equation}
where
\begin{displaymath}
f(x) = e^{- \lambda^2 \left( x - N / 2 \right)^2}.
\end{displaymath}
\smallskip

For compactness we %define $\phi_N = e^{- \lambda^2 N^2 / 4}$ and
let $S_d = e^{- \lambda^2 N^2 / 4} \ PG_d(U)$. We
now bound $S_d$ using the following version of the Euler-Maclaurin summation formula, truncated after
the term involving the first derivatives (see \cite[p.211]{FROB65}):
\begin{equation}\label{eq:em1}
S_d = \sum_{i=0}^d f(i) = \int\limits_0^d f(x) \mathrm{d} x + \frac{1}{2} \left[ f(0) + f(d)
\right] + \frac{1}{12} \left[ f'(d) - f'(0) \right] - R_d,
\end{equation}
where the remainder term $R_d$ satisfies
\begin{displaymath}
R_d = \frac{d}{720} \ f^{(4)} (\xi),
\end{displaymath}
for some $\xi$, with $0 < \xi < d$.
\smallskip

We first consider the definite integral. Making the substitution
\begin{displaymath}
y = \lambda (x - N/2),
\end{displaymath}
%and letting $d = \Delta = \lfloor N \rfloor$,
we obtain
\begin{displaymath}\label{eq:em2}
\int\limits_0^d f(x) \ \mathrm{d} x = \frac{1}{\lambda} \ \int\limits_{- \lambda
N/2}^{\lambda n/2} e^{- y^2} \ \mathrm{d} y,
\end{displaymath}
where $n = 2 d - N$.

%where $n = 2 d - N = 2 \lfloor N \rfloor - N$, so $N-2 < n \le N$.

\smallskip

Expressing this in terms of the well-known {\em error function} \cite[7.1.1]{ABRA72},
\begin{displaymath}
{\rm erf}(x) = \frac{2}{\sqrt{\pi}} \int\limits_0^x e^{-y^2} \ \mathrm{d} y,
\end{displaymath}
and using that fact that $e^{-y^2}$ is an even function, we obtain
\begin{equation}\label{eq:erf}
\int\limits_0^d f(x) \ \mathrm{d} x = \frac{\sqrt{\pi}}{2 \lambda} \ \left( {\rm erf}
\left( \frac{\lambda N}{2} \right) + {\rm erf}\left( \frac{\lambda n}{2} \right) \right).
\end{equation}
\medskip

Using the formulae in the Appendix to get an expression for $f'$, we easily obtain the following expression
for the other terms on the right-hand side of (\ref{eq:em1}), apart from the remainder term $R_d$,
\begin{equation}\label{eq:terms}
\frac{1}{2} \left[ f(0) + f(d)\right] + \frac{1}{12} \left[ f'(d) - f'(0) \right] = \left(
\frac{1}{2} - \frac{\lambda^2 N}{12} \right)  e^{- \lambda^2 N^2 / 4}  + \left( \frac{1}{2} -
\frac{\lambda^2 n}{12} \right)   e^{- \lambda^2 n^2 / 4} .
\end{equation}
\smallskip

We now turn our attention to the remainder term $R_d$. This satisfies
\begin{displaymath}
R_d = \frac{d}{720} \ f^{(4)} (\xi) = \frac{\lambda^4 d}{720} \ F^{(4)} (\eta),
\end{displaymath}
for some $\eta$, with $- \lambda N/2 < \eta < \lambda n/2$, where $F(\eta) = e^{-\eta^2}$.

\smallskip

Using (\ref{eq:ap}) in the Appendix, this gives
\begin{equation}\label{eq:rem}
- \frac{\lambda^4 d}{96} < R_d \le \frac{\lambda^4 d}{60}.
\end{equation}

Substituting (\ref{eq:erf}) and (\ref{eq:terms}) into (\ref{eq:em1}), we obtain
\begin{equation}\label{eq:epsilon}
S_d = \frac{\sqrt{\pi}}{2 \lambda} \ \left( {\rm erf} \left( \frac{\lambda N}{2}
\right) + {\rm erf}\left( \frac{\lambda n}{2} \right) \right) + \left( \frac{1}{2} -
\frac{\lambda^2 N}{12} \right)  e^{- \lambda^2 N^2 / 4} + \left( \frac{1}{2} - \frac{\lambda^2 n}{12}
\right)  e^{- \lambda^2 n^2 / 4} - R_d .
\end{equation}
\smallskip

Together with (\ref{eq:rem}), this immediately gives bounds on $PG_d(U)$ since
$PG_d(U) = e^{\lambda^2 N^2 / 4} S_d$. We may then estimate the total potential gain $PG(U)$ as $PG_\Delta(U)$ by putting $d = \Delta$.

\section{Distribution of Potential Gain Values}
\label{sec:analysis}

In this section we examine some aspects of the potential gain function and the distribution of its values.

\medskip

\begin{figure}[ht]
\centerline{\includegraphics[width=12cm,height=9.33cm]{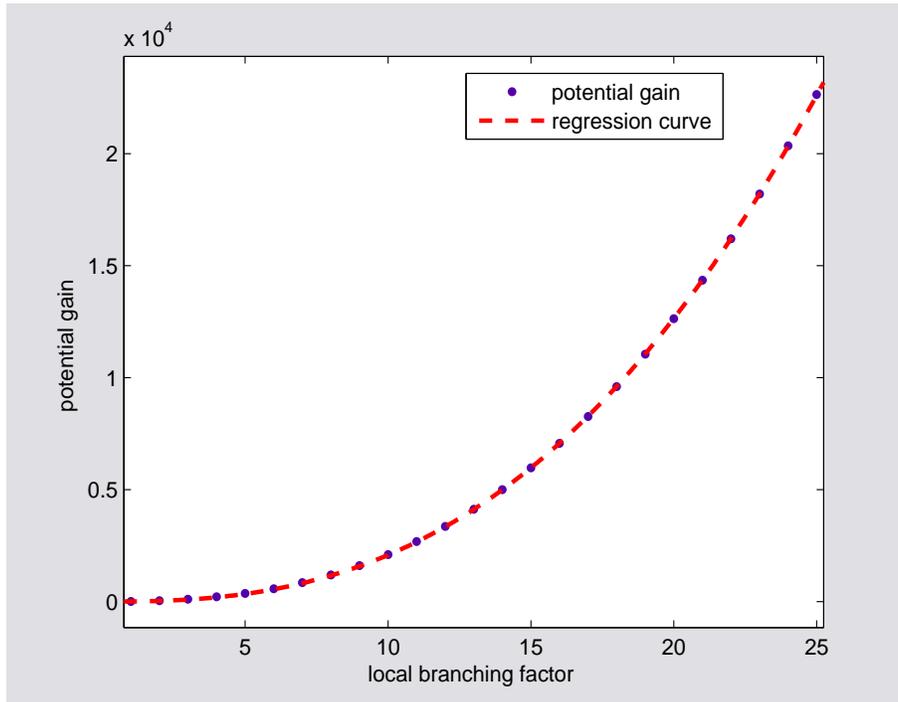}}
\caption{\label{fig:pg-bk} Plot of $PG_{10}(U)$ against $\beta$}
\end{figure}

We assume that $\Delta$, the mean number of user clicks per navigation session, is about 10;
this is quite close to 8.32 reported in \cite{HUBE98}. We also assume that
the local branching factor $\beta$ is between $2$ and $25$; see \cite{DILL02} for
data on branching factors for different subsets of the web. We note that, in the case
when $\beta = 1$, we have $N = \Delta = 1$ and $PG_{10}(U) = 2$, implying that there is no
choice for the user. In Table~\ref{table:delta} we give, for $2 \le \beta \le 25$, various quantities related to the potential gain. These were computed as follows:
\renewcommand{\labelenumi}{(\roman{enumi})}
\begin{enumerate}
\item From (\ref{eq:N}), $\delta = \beta^{- 2/(N-1)}$.

\item By definition, $\lambda = (\frac{1}{2} \ln (1/\delta))^{1/2}$.

\item From (\ref{eq:max-total-leaves}) we know that {\em max}, the maximum of $a^x \theta^{x^2}$, is attained for $x = N/2$; the maximum value is easily shown to be $\beta^{N^2 /4(N-1)}$.

\item $PG_{10}(U)$,  our estimate of $PG(U)$, is given by (\ref{eq:pg}). By
regression on the plot shown in Figure~\ref{fig:pg-bk}, we found that this is a good fit to the
power-law:
\begin{displaymath}
PG_{10}(U) \approx 5.162 \, \beta^{2.605}.
\end{displaymath}

\item {\em noR} is the approximation to $PG_{10}(U)$ obtained from (\ref{eq:epsilon}) if we set $R_{10}$ to $0$.

\item The upper ({\em ub}) and lower ({\em lb}) bounds on $PG_{10}(U)$ are derived
from (\ref{eq:rem}) and (\ref{eq:epsilon}).

\end{enumerate}
\smallskip

\begin{figure}[ht]
\centerline{\includegraphics[width=12cm,height=9.33cm]{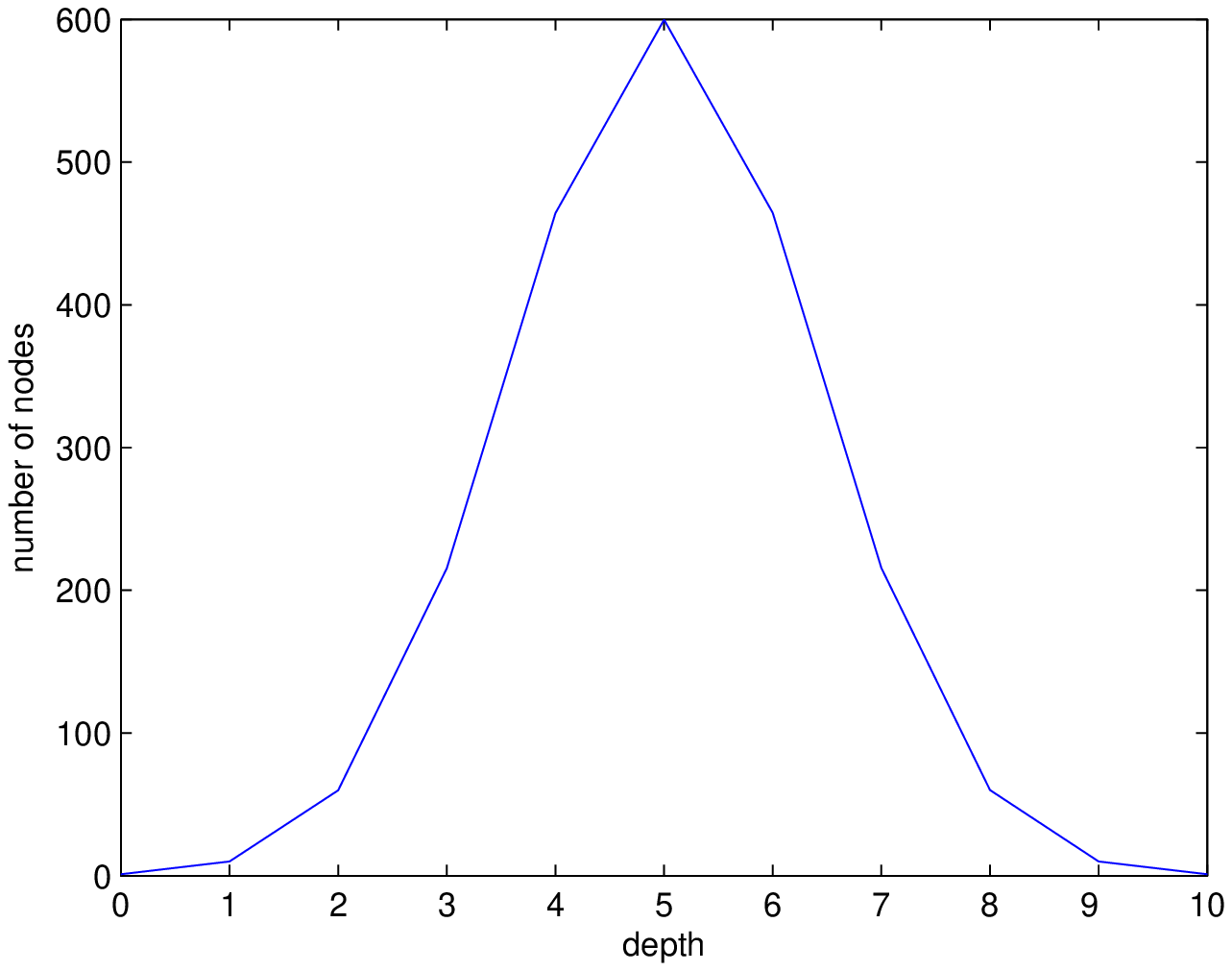}}
\caption{\label{fig:pg-10} Plot of the distribution of $PG_d(U)$ for $N = 10$ and $\beta =
10$}
\end{figure}

Although the average of the lower and upper bounds given in Table~\ref{table:delta} yields a reasonably good approximation to $PG_{10}(U)$, we see that {\em noR}, i.e. the approximation obtained if we ignore the remainder term, is extremely close to the actual value of $PG_{10}(U)$. In Figure~\ref{fig:pg-10} we show a typical plot of the potential gain against the depth $d$.

\begin{table}[ht]
\begin{center}
\begin{tabular}{|l|l|l|l|l|l|l|l|l|}
\hline $\beta$ & $\delta$ & $\lambda$ & $max$ & $PG_{10}(U)$ & $noR$ &$lb$ & $ub$ &
$(lb+ub)/2$ \\ \hline \hline 2&0.86&0.28&6.86&42.49&42.49&42.49&42.5&42.49\\\hline
3&0.78&0.35&21.15&106.65&106.65&106.6&106.68&106.64\\\hline
4&0.73&0.39&47.03&211.98&211.97&211.79&212.09&211.94\\\hline
5&0.7&0.42&87.41&366.08&366.07&365.6&366.36&365.98\\\hline
6&0.67&0.45&145.05&575.98&575.96&575&576.56&575.78\\\hline
7&0.65&0.46&222.58&848.26&848.24&846.51&849.32&847.91\\\hline
8&0.63&0.48&322.54&1189.17&1189.15&1186.28&1190.94&1188.61\\\hline
9&0.61&0.49&447.38&1604.7&1604.67&1600.23&1607.45&1603.84\\\hline
10&0.6&0.51&599.48&2100.59&2100.55&2094.01&2104.64&2099.33\\\hline
11&0.59&0.52&781.19&2682.38&2682.34&2673.1&2688.12&2680.61\\\hline
12&0.58&0.53&994.78&3355.48&3355.43&3342.79&3363.33&3353.06\\\hline
13&0.57&0.53&1242.47&4125.1&4125.05&4108.23&4135.57&4121.9\\\hline
14&0.56&0.54&1526.47&4996.36&4996.31&4974.43&5009.98&4992.21\\\hline
15&0.55&0.55&1848.93&5974.24&5974.18&5946.29&5991.62&5968.95\\\hline
16&0.54&0.56&2211.96&7063.61&7063.56&7028.57&7085.42&7057\\\hline
17&0.53&0.56&2617.66&8269.26&8269.2&8225.97&8296.22&8261.1\\\hline
18&0.53&0.57&3068.09&9595.88&9595.81&9543.07&9628.78&9585.92\\\hline
19&0.52&0.57&3565.28&11048.06&11047.99&10984.39&11087.74&11036.07\\\hline
20&0.51&0.58&4111.23&12630.34&12630.27&12554.35&12677.72&12616.03\\\hline
21&0.51&0.58&4707.94&14347.18&14347.1&14257.31&14403.22&14330.27\\\hline
22&0.5&0.59&5357.37&16202.96&16202.88&16097.56&16268.71&16183.13\\\hline
23&0.5&0.59&6061.46&18202.01&18201.93&18079.31&18278.57&18178.94\\\hline
24&0.49&0.59&6822.13&20348.61&20348.53&20206.75&20437.14&20321.94\\\hline
25&0.49&0.6&7641.29&22646.97&22646.88&22483.97&22748.69&22616.33\\\hline
\end{tabular}
\end{center}
\caption{\label{table:delta} Tabulation for $N = \Delta = 10$}
\end{table}
\medskip

\medskip

\medskip

\section{An Alternative Discounting Factor}
\label{sec:poisson}

We next look at an alternative discounting factor. We assume that, at distance $d$ from the
starting URL $U$, the user will only inspect $1/(d +1)$ of the currently available
links, prior to following one of them. Thus, given $\beta$, the effective branching
factor at depth $i$ is $\beta / (i+1)$ and the potential number of available nodes nodes at
this depth is
\begin{equation}\label{eq:alt-pg-depthi}
\frac{\beta}{1} \ \frac{\beta}{2} \cdots \frac{\beta}{i} =
\frac{\beta^i}{i !},
\end{equation}
which corresponds to (\ref{eq:pg-depthi}).

\medskip

The {\em alternative potential gain} of $U$, denoted by $APG(U)$, is now simply the
total number of available nodes at all depths, i.e.
\begin{equation}\label{eq:alt-pg-define}
APG(U) = \sum_{i=0}^\infty \frac{\beta^i}{i!} = e^{\beta},
\end{equation}
which corresponds to (\ref{eq:pg-define}). The alternative potential gain of $U$ up to depth $d$, denoted by $APG_d(U)$, is obtained by replacing the upper limit of the sum by $d$.

\medskip

As in the case of $PG(U)$, in order to approximate $APG(U)$, we need to find the
maximum depth $d$ such that
\begin{displaymath}
\frac{\beta^d}{d !} \ge 1.
\end{displaymath}
\smallskip

By Stirling's approximation \cite{GRAH94}, $\ln (d !) \approx d \ln d - d$, so we have (approximately) that
$d \le e \beta$. Thus $\Delta = \lfloor e \beta \rfloor$.

\medskip

We next consider the maximum term in the sum (\ref{eq:alt-pg-define}), i.e.
$APG(U)$, as we did for $PG(U)$. It is straightforward to show that the maximum term is at
$i = \lfloor \beta \rfloor$. Ignoring rounding errors and using Stirling's approximation, we see that the maximum term is approximately $e^{\beta} / (2 \pi {\beta})^{1/2}$. (When $\beta$ is an integer the maximum is also attained when $i = \beta - 1$.)

\medskip

\begin{figure}[ht]
\centerline{\includegraphics[width=12cm,height=9.33cm]{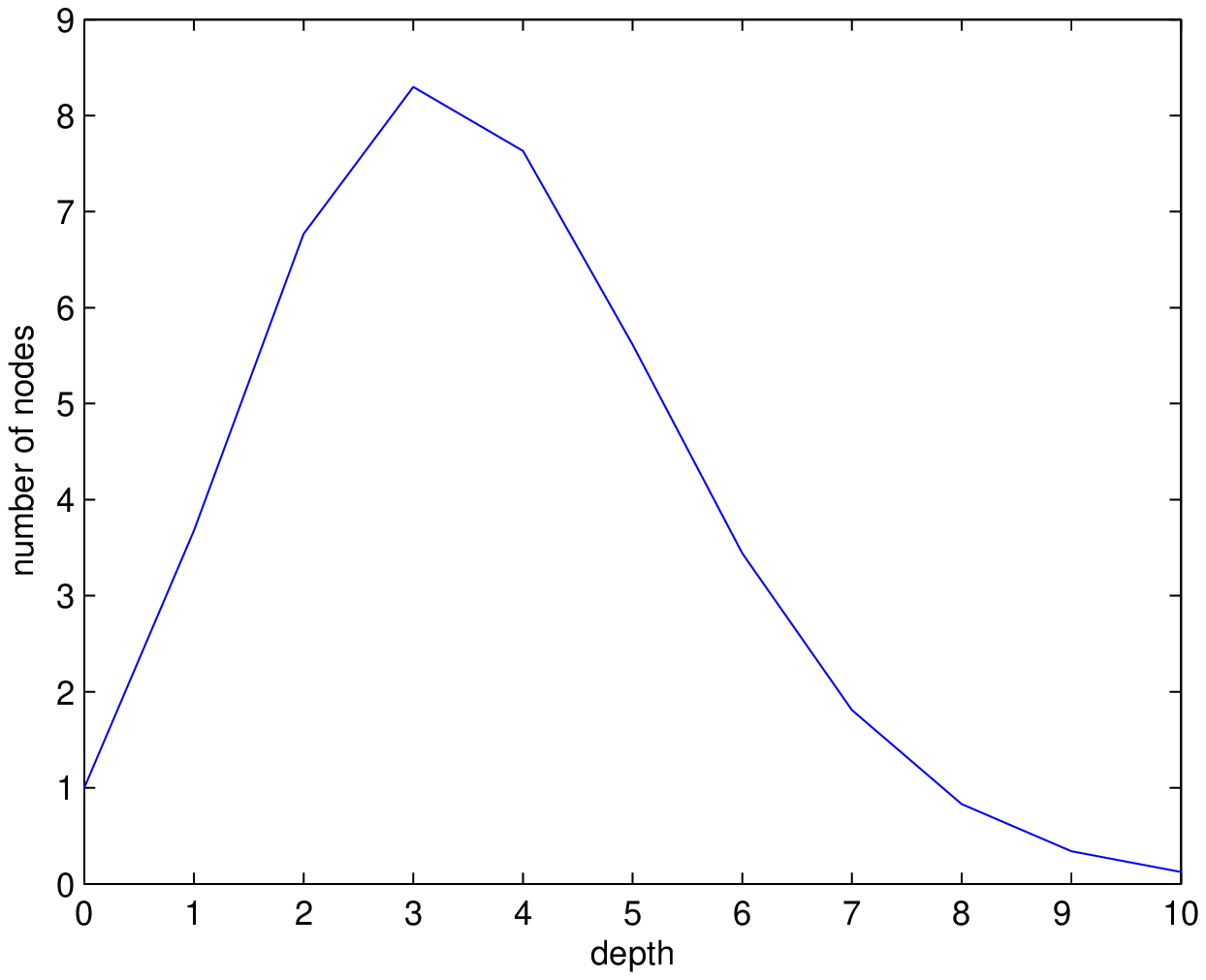}}
\caption{\label{fig:pg-poisson-10} Plot of the distribution of  $APG_d(U)$ for $\Delta=10$ and $\beta = 3.6788$}
\end{figure}

Taking $d = \Delta = 10$ we obtain the branching factor $\beta = 3.6788$.
In Figure~\ref{fig:pg-poisson-10} we show a  typical plot of the alternative potential gain against the depth $d$. In this case the sum of the first $10$ terms in (\ref{eq:alt-pg-define}) is $39.54$, which is a good approximation of $e^{3.6788} = 39.60$.

\medskip

\section{Concluding Remarks}
\label{sec:concluding}

We defined a measure of navigability, called the potential gain, that provides a model of user
navigation in the web. This can help the user in an online scenario to choose a starting URL for navigation, given no other information. One important factor that distinguishes the potential gain from other link
analysis metrics \cite{LANG06}, such as Google's PageRank, is that it measures ``hubness'', i.e. the
accessibility from the page of information on others pages, rather than authority, i.e. the accessibility from elsewhere of information on the page. (See also our comment after (\ref{eq:pg-define}) regarding another important distinction between the potential gain and PageRank.) Whereas PageRank measures authority, the Hyperlink-Induced Topic Search (HITS) algorithm \cite{KLEI99b} identifies both hubs and authorities, but its computation is query-specific. In this context, it is worth noting that the potential gain is related to the notion of centrality \cite{FREE79}, which is a fundamental notion in social network analysis \cite{SCOT00}.

\smallskip

The potential gain has been applied in a search and navigation engine that we have
developed. Its distinctive feature is that an answer to a user query suggests several possible
navigation paths that the user can follow \cite{WHEE03}, rather than just individual web pages
as suggested by conventional search engines. As part of the search and navigation engine, potential gain values are pre-computed for each page in the web site being searched; these are then used to
select good starting URLs for navigation \cite{LEVE04a}.

\section*{Appendix}

We obtain here the derivatives of $f(x)$. To simplify the calculations, it is convenient to
let $y = \lambda (x - N/2)$ and define
\begin{displaymath}
F(y) = e^{- y^2} = f(x).
\end{displaymath}

\noindent The derivatives of $f(x)$ are determined from the derivatives of $F(y)$, since
\begin{displaymath}
f^{(k)} (x) = \lambda^{k} F^{(k)} (y).
\end{displaymath}

By straightforward differentiation we obtain:
\begin{eqnarray*}
F'(y) & = & - 2 y \ e^{-y^2} \\
F''(y)&  = & (4 y^2 - 2) \ e^{-y^2} \\
F'''(y) & = & (- 8 y^3 + 12 y) \ e^{-y^2} \\
F^{(4)}(y) &  = & (16 y^4 - 48 y^2 + 12) \ e^{- y^2} \\
F^{(5)}(y) & = &  (-32 y^5 + 160 y^3 - 120 y) \ e^{- y^2} \\
\end{eqnarray*}

These functions are closely related to the Hermite polynomials \cite[p.189]{FROB65}.

\smallskip

We also require the extreme values of $F^{(4)}(y)$. Using a straightforward calculation, it is
readily verified that the local extrema of this function are
\begin{eqnarray*}
F^{(4)}(y) & =  & 12 \quad \mathrm{at} \  y = 0, \\
F^{(4)}(y) & =  & - 7.42 \quad \mathrm{at} \ y = \pm (( 5 - \sqrt{10} ) /2)^{1/2}, \\
F^{(4)}(y) & =  & 1.39 \quad \mathrm{at} \ y = \pm (( 5 + \sqrt{10} ) /2)^{1/2}.
\end{eqnarray*}

Thus
\begin{equation}\label{eq:ap}
-7.5 < F^{(4)}(y) \le 12.
\end{equation}

\newcommand{\etalchar}[1]{$^{#1}$}

\end{document}